\documentclass[aps,twocolumn,showpacs,preprintnumbers,amsmath,amssymb]{revtex4}

\usepackage{txfonts}
\usepackage{dcolumn}
\usepackage{mathrsfs}
\usepackage{bm}
\usepackage{amsmath,amssymb,epsfig,float}

\begin{document}

\title{Quantum decoherence in noninertial frames}

\author{Jieci Wang and Jiliang Jing\footnote{Corresponding author,
 Email: jljing@hunnu.edu.cn}}
\affiliation{ Institute of Physics and  Department of  Physics,
\\ Hunan Normal University, Changsha, \\ Hunan 410081, P. R. China
\\ and
\\ Key Laboratory of Low-dimensional Quantum Structures
\\ and Quantum
Control of Ministry of Education, \\ Hunan Normal University,
Changsha, Hunan 410081, P. R. China}

\vspace*{0.2cm}
\begin{abstract}
\vspace*{0.2cm} Quantum decoherence, which appears when a system
interacts with its environment in an irreversible way, plays a
fundamental role in the description of quantum-to-classical
transitions and has been successfully applied in some important
experiments. Here, we study the decoherence in noninertial frames
for the first time. It is shown that the decoherence and loss of the
entanglement generated by the Unruh effect will influence each other
remarkably. It is interesting to note that in the case of the total
system under decoherence, the sudden death of entanglement may
appear for any acceleration. However, in the case of only Rob's
qubit underging decoherence sudden death  may only occur when the
acceleration parameter is greater than a ``critical point."
\end{abstract}

\vspace*{1.5cm}
 \pacs{03.65.Ud, 03.67.Mn, 04.70.Dy,  97.60.Lf}

\maketitle

\section{introduction}

The study of quantum information in noninertial framework is not
only helpful for understanding some key questions in quantum
mechanics \cite{Peres,Boschi,Bouwmeester}, but it also plays an
important role in the study of entropy and the information paradox
of black holes \cite{Bombelli-Callen, Hawking-Terashima}. Recently,
much attention has been focused on the topic of the quantum
information in a relativistic setting \cite{SRQIT1,Ging,
Alsing-Mann,Qiyuan,jieci1,jieci2,Lamata} and, in particular, on how
the Unruh effect changes the degree of quantum entanglement
\cite{Schuller-Mann} and fidelity of teleportation
\cite{Alsing-Milburn}. However, it should be pointed out that all
investigations in noninertial frames are confined to the studies of
the quantum information in { \em an isolated system}. However, in a
realistic quantum system, the {\em interaction} between the quantum
system and  the surrounding environment is inevitable, and then the
dynamics of the system is non-unitary (although the combined system
plus environment evolves in a unitary fashion). The decoherence
\cite{Zurek,Breuer}, which appears  when a system interacts with its
environment in a irreversible way, can be viewed as the transfer of
information from system into the environment. It plays a fundamental
role in the description of the quantum-to-classical transition
\cite{Giulini, Schlosshauer} and has been successfully applied in
the cavity QED \cite{Brune} and ion trap experiments \cite{Myatt}.

In this article we investigate the quantum decoherence of Dirac
fields in a noninertial system. For the sake of brevity and without
loss of generality, we  consider only the amplitude damping channel
\cite{Salles},which is the most typical quantum noisy channel and
can be modeled by the spontaneous decay of a two-level quantum state
in an electromagnetic field \cite{Brune1}. We assume that two
observers, Alice and Rob, share an entangled initial state at the
same point in flat Minkowski spacetime. After that Alice stays
stationary while Rob moves with uniform acceleration. We let one (or
both) of the observers moves (or stays) in the noisy environment and
discuss whether or not the quantum decoherence and the loss of
entanglement generated by Unruh radiation will influence each other.
A key question to be answered is: Does r the entanglement appears to
be sudden death \cite{Yu} or does it only disappears as time tends
to infinity?


\vspace*{0.5cm}

\begin{figure}[ht]
\includegraphics[scale=0.7]{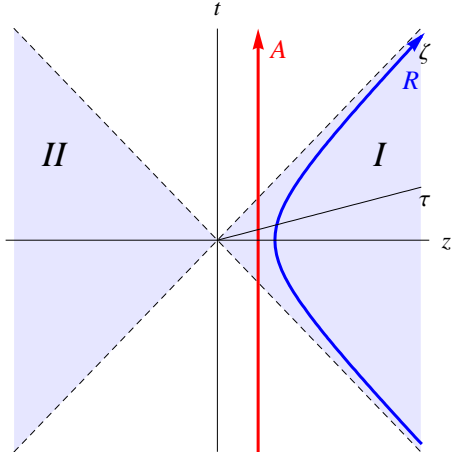}
\caption{\label{Rindler}(Color online) Rindler spacetime diagram: An
accelerated observer Rob travels on a hyperbola in region $I$ with
uniform acceleration $a$ and is causally disconnected from region
$II$.}
\end{figure}

We assume that Alice has a detector sensitive  only to mode
$|n\rangle_{A}$ and Rob has a detector sensitive only to mode
$|n\rangle_{R}$, and they share the maximally entangled initial
state
\begin{eqnarray}\label{initial}
|\Phi\rangle_{AR}=\frac{1}{\sqrt{2}}(|0\rangle_{A}|0\rangle_{R}
+|1\rangle_{A}|1\rangle_{R}),
\end{eqnarray} at the
same point in Minkowski spacetime, where $\{|n\rangle_{A}\}$ and
$\{|n\rangle_{R}\}$ indicate Minkowski modes described by Alice and
Rob, respectively. We then let Alice remain stationary while Rob
moves with uniform acceleration. From the perspective of Rob the
Minkowski vacuum is found to be a two-mode squeezed state
\cite{Alsing-Mann}
\begin{eqnarray}\label{Dirac-vacuum}
|0\rangle_{M}= \cos r|0\rangle_{I}|0\rangle _{II}+\sin
r|1\rangle_{I}|1\rangle _{II},
\end{eqnarray}
where $\cos r=(e^{-2\pi\omega c/a}+1)^{-1/2}$, $a$ is Rob's
acceleration, $\omega$ is frequency of the Dirac particle, $c$ is
the speed of light in vacuum, and  $\{|n\rangle_{I}\}$ and
$\{|n\rangle_{II}\}$ indicate Rindler modes in Region $I$ and $II$
(see Fig. \ref{Rindler}), respectively. The only excited state is
given by
\begin{eqnarray}\label{Dirac-excited}
|1\rangle_{M}=|1\rangle_{I}|0\rangle_{II}.
\end{eqnarray}

Using Eqs. (\ref{Dirac-vacuum}) and (\ref{Dirac-excited}), we can
rewrite Eq. (\ref{initial}) in  terms of Minkowski modes for
Alice and  Rindler modes for Rob
\begin{eqnarray} \label{state}
|\Phi\rangle_{A,I,II}&=&\frac{1}{\sqrt{2}}\bigg( \cos r|0\rangle_{A}
|0\rangle_{I}|0\rangle_{II}+\sin r|0\rangle_{A}
|1\rangle_{I}|1\rangle_{II}\nonumber \\&&+|1\rangle_{A}
|1\rangle_{I}|0\rangle_{II}\bigg).
\end{eqnarray}
Since Rob is causally disconnected from region $II$, the physically
accessible information is encoded in the mode $A$ described by Alice
and  mode $I$ described by Rob. Tracing over the state in region
$II$, we obtain
\begin{eqnarray}  \label{eq:state1}
\rho_{A,I}&=&\frac{1}{2}\bigg[\cos^2 r|00\rangle\langle00|+\cos
r(|00\rangle\langle11|+|11\rangle\langle00|)\nonumber \\&&+\sin^2
r|01\rangle\langle01|+|11\rangle\langle11|\bigg],
\end{eqnarray}
where $|mn\rangle=|m\rangle_{A}|n\rangle_{I}$.

 \section{ Case of Single qubit undergoing decoherence}

\vspace*{0.5cm}

{\it Single qubit under decoherence case:}  Now we consider Rob's
state coupled to a dissipative environment, which corresponds to the
spontaneous decay of Rob's state because it interacts with an
electromagnetic field environment \cite{Brune1}. This process may be
described as \cite{Breuer}
\begin{eqnarray}
\label{AmplitudeDampingMap}
|0\rangle_{R}|0\rangle_E&\rightarrow&
|0\rangle_{R}|0\rangle_E  \label{en1}\;,\\
|1\rangle_{R}|0\rangle_E&\rightarrow&
\sqrt{1-P_{R}}|1\rangle_{R}|0\rangle_E +
\sqrt{P_{R}}|0\rangle_{R}|1\rangle_E  \label{en2}\;.
\end{eqnarray}
Eq. (\ref{en1}) indicates that the system has no decay and the
environment is untouched. Eq. (\ref{en2}) shows that, if decay
exists in the system, it can either remain there with probability
$(1-P_{R})$, or be transferred into the environment with probability
$P_{R}$. Usually, the dynamic of an open quantum system is described
by a reduced density operator which is obtained from the density
operator of the total system by tracing over the degrees of freedom
of the environment. By considering the environment as a third
system, we can obtain a unified entanglement-only picture.

The dynamics described by Eqs. ($\ref{en1}$) and ($\ref{en2}$) for a
single qubit also can be represented by the following Kraus
operators \cite{Kraus,Choi}
 \begin{eqnarray}
M_0^{R}= \left(\begin{array}{cc}
           1&0\\
           0&\sqrt{1-P_{R}}
           \end{array}\right),&\;& M^{R}_1=\left(\begin{array}{cc}
                                           0&\sqrt{P_{R}}\\
                                           0&0
                                          \end{array}\right),
\label{Kraus1B}
\end{eqnarray}
where $P_{R}$ ($0\leq P_{R}\leq1$) is a parameter relating only to
time. Under the Markov approximation, the relationship between the
parameter $P_{R}$ and the time $t$ is given by $P_{R}=(1-e^{-\Gamma
t})$ \cite{Brune1,Salles} where $\Gamma$ is the decay rate.

As a first step toward the study of quantum decoherence, we rewrite
the state Eq. (\ref{eq:state1}) as
\begin{eqnarray}  \label{eq:state2}
\rho_{A,I}&=&\frac{1}{2}\bigg[|0\rangle_A\langle0|\otimes
\mathrm{T}^{00}_{R}+ |0\rangle_A\langle1|\otimes
\mathrm{T}^{01}_{R}\nonumber \\&& +|1\rangle_A\langle0|\otimes
\mathrm{T}^{10}_{R}+ |1\rangle_A\langle1|\otimes
\mathrm{T}^{11}_{R}\bigg],
\end{eqnarray}
with
\begin{eqnarray}
\nonumber && \mathrm{T}^{00}_{R}=\left(\begin{array}{cc}
           \cos^2 r&0\\
           0&\sin^2 r
           \end{array}\right), ~\;
           \mathrm{T}^{01}_{R}=\left(\begin{array}{cc}
                                           0& 0\\
                                          \cos r& 0
                                          \end{array}\right),
\\
\nonumber &&  \mathrm{T}^{10}_{R}=\left(\begin{array}{cc}
           0&\cos r\\
           0&0
           \end{array}\right), ~~~~~~~~~\;
           \mathrm{T}^{11}_{R}=\left(\begin{array}{cc}
                                           0&0\\
                                           0&1
                                          \end{array}\right).
\label{Kraus2B}
\end{eqnarray}
This form of the state suggests a natural bipartite split. We can
use it to study how the environment effects Rob's single qubit.
Under the amplitude damping channel, the state evolves to
 \begin{eqnarray}\label{eq:state3}
 \rho_{s}=\frac{1}{2}\left(
  \begin{array}{cccc}
    1-\beta \sin^2 r & 0 & 0 & \sqrt{\beta} \cos r \\
    0 & \beta \sin^2 r & 0 & 0 \\
    0 & 0 & P_{R} & 0 \\
    \sqrt{\beta} \cos r & 0 & 0 & \beta \\
  \end{array}
\right),
\end{eqnarray}
where $\beta=1-P_R$.

It is well known that the degree of entanglement for two-qubits
mixed state in noisy environments can be quantified conveniently by
concurrence, which is defined as \cite{Wootters,Coffman}
\begin{eqnarray}  \label{Concurrence}
C_{s} =\max \left\{ 0,\sqrt{\lambda _{1}}-\sqrt{\lambda
_{2}}-\sqrt{\lambda _{3}}-\sqrt{\lambda _{4}}\right\}, \quad\lambda_i\ge
\lambda_{i+1}\ge 0,
\end{eqnarray}
where $\sqrt{\lambda_i}$ are square root of the eigenvalues of the
matrix $\rho_{s}\tilde{\rho}_{s}$, where
$\tilde{\rho}_{s}=(\sigma_y\otimes\sigma_y)\,
\rho_{s}^{*}\,(\sigma_y\otimes\sigma_y)$ is the ``spin-flip" matrix
for the state (\ref{eq:state3}) which is given by
\begin{eqnarray}\label{eq:state4}
 \tilde{\rho}_{s}=\frac{1}{2}\left(
  \begin{array}{cccc}
    \beta & 0 & 0 & \sqrt{\beta} \cos r \\
    0 &  P_{R} & 0 & 0 \\
    0 & 0 & \beta \sin^2 r & 0 \\
    \sqrt{\beta} \cos r & 0 & 0 & 1-\beta \sin^2 r \\
  \end{array}
\right).
\end{eqnarray}
Hence, the eigenvalues of $\rho_{s}\tilde{\rho}_{s}$ are
\begin{eqnarray}
\nonumber
&&\lambda_1=\frac{\beta}{4}\bigg[\cos^2 r
+\bigg(\cos r+\sqrt{\cos^2 r+P_{R}\sin^2 r}
\bigg)^2\bigg],\\ \nonumber
&&\lambda_2=\frac{\beta}{4}\bigg[\cos^2 r+
\bigg(\cos r-\sqrt{\cos^2 r+P_{R}\sin^2 r}\bigg)^2\bigg],\\
&&\lambda_3=\lambda_4=\frac{\beta}{4}P_{R}\sin^2 r.
\end{eqnarray}
By using Eq. (\ref{Concurrence}) we get the concurrence which is
$\cos r$ when the decay parameter $P_{R}=0$,  in which case our
result reverts to that of Ref. \cite{Alsing-Mann}.

\begin{figure}[ht]
\includegraphics[scale=0.75]{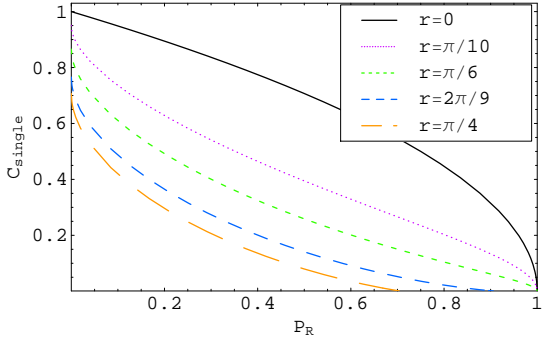}
\caption{\label{ERP1}(Color online) Concurrence as a functions of
the decay parameter $P_{R}$  with some fixed acceleration parameters
[$r=0$ (black line), ~$\frac{\pi}{10}$ (dotted line),
~$\frac{\pi}{6}$ (dashed green line), ~$\frac{2\pi}{9}$ (dashed blue
line), ~$\frac{\pi}{4}$ (dashed orange line)] when only Rob's qubit
undergoes decoherence.}
\end{figure}

In Fig. (\ref{ERP1}) we plot the behavior of the concurrence which
shows how the acceleration of Rob would change the properties of
entanglement when his qubit couples to the environment. It is shown
that, compared with the case of $P_{R}=0$ \cite{Alsing-Mann}
(isolated system), the degree of entanglement decreases rapidly as
acceleration increases. It is worth to note that Alsing {\it et al}
\cite{Alsing-Mann} found that the entanglement of Dirac fields in an
isolated system is not completely destroyed even in the limit case
that Rob is under infinite acceleration. But we find that the
entanglement of Dirac fields could tend to zero for  finite
acceleration. That is to say, the noise can greatly influence the
loss of the entanglement generated by Unruh effect. Note that
$P_{R}$ is a monotonically increasing function of the time, this
figure in fact describes the time evolution of entanglement of a
bipartite system when one of them is coupled to an amplitude damping
environment. It is interesting to note that the entanglement only
disappears as $t \rightarrow \infty$ when the acceleration is small
or zero. However, the sudden death of entanglement appears at a
finite time for large and infinite accelerations. Obviously, in the
time evolution of entanglement there exists a ``critical point" for
the acceleration parameter. We note that the concurrence $C_{s}=0$
if the acceleration parameter $r$ and the decay parameter $P_R$
satisfy the relation
\begin{eqnarray}
r=\arcsin \left(\frac{\sqrt{P_R^2+4}-P_R}{2}\right).
\end{eqnarray}
Considering the condition $0\leq P_{R}\leq 1$, we find  that sudden
death of the entanglement will appear when $\arcsin
[(\sqrt{5}-1)/2]\leq r\leq \frac{\pi}{4}$. Thus, the ``critical
point" is $r_c=\arcsin [(\sqrt{5}-1)/2]=0.666239$ below which sudden
death of the entanglement can not take place.

\section{Case of two qubits undergoing decoherence }

\vspace*{0.5cm}

{\it Two qubits under decoherence case:} Now we consider both Alice
and Rob's states coupled to the noisy environment, which acts
independently on both their states. The total evolution of this two
qubits system  can be expressed as
\begin{eqnarray}
L(\rho_{AR})=\sum_{\mu \nu} M^{A}_\mu \otimes M^{R}_\nu \rho_{AR}
M_\nu^{R\dag}\otimes M_\mu^{A\dag}, \label{EvolKraus}
 \end{eqnarray}
where $M_{\mu}^{i}$ are the Kraus operators
\begin{eqnarray}
M_0^{i}=\left(\begin{array}{cc}
           1&0\\
           0&\sqrt{1-P_{i}}
           \end{array}\right), &\;& M^{i}_1=\left(\begin{array}{cc}
                                           0&\sqrt{P_{i}}\\
                                           0&0
                                          \end{array}\right),
\label{Kraus1A}
\end{eqnarray}
where $i=(A,~R)$, $P_{A}$  is the decay parameter in Alice's quantum
channel and $P_{R}$  is Rob's decay parameter.  Here we only
consider the global channels \cite{Salles}, in which all the
subsystems are embedded in the same environment (i.e.,
$P_{A}=P_R=P$).

When both of the two qubits are coupled to the environment, state
Eq. (\ref{eq:state1}) evolves to
 \begin{eqnarray}\label{eq:state5}
 &&\rho_{t}=\frac{1}{2}\left(
  \begin{array}{cccc}
    1+P^2-\tilde{\beta} \sin^2 r & 0 & 0 & \tilde{\beta} \cos r \\
    0 & \tilde{\beta} (P+\sin^2 r) & 0 & 0 \\
    0 & 0 & P \tilde{\beta} & 0 \\
    \tilde{\beta} \cos r & 0 & 0 & \tilde{\beta}^2 \\
  \end{array}
\right),\nonumber \\
\end{eqnarray}
where $\tilde{\beta}=1-P$. We can easily get the ``spin-flip" of
this state and find that the matrix $\rho_{t}\tilde{\rho}_{t}$ has
eigenvalues
\begin{eqnarray}
\nonumber &&\tilde{\lambda}_1=\frac{\tilde{\beta}^2}{4}
\bigg[\cos^2 r+\bigg(\cos r+\sqrt{1+P^2-\tilde{\beta}
\sin^2 r}\bigg)^2\bigg],\\ \nonumber\
&&\tilde{\lambda}_2=\frac{\tilde{\beta}^2}{4}\bigg[\cos^2 r
+\bigg(\cos r-\sqrt{1+P^2-\tilde{\beta} \sin^2 r}\bigg)^2\bigg],\\
&&\tilde{\lambda}_3=\tilde{\lambda}_4=\frac{\tilde{\beta}^2}{4}
P(P+\sin^2 r).
\end{eqnarray}
It is interesting to note that the concurrence is also $\cos r$ for
$P=0$.

\begin{figure}[ht]
\includegraphics[scale=0.75]{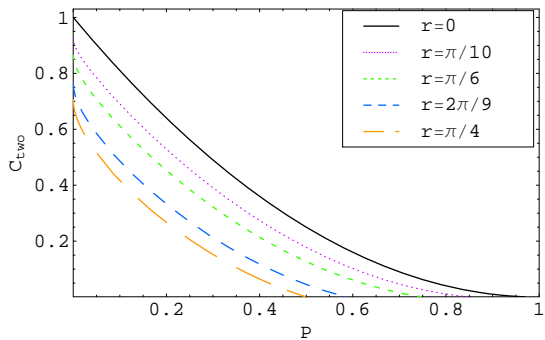}
\caption{\label{ERP2}(Color online) The concurrence as functions of
the decay parameter $P$ and acceleration parameter $r$ [$r=0$ (black line), ~$\frac{\pi}{10}$ (dotted line), ~$\frac{\pi}{6}$ (dashed green line), ~$\frac{2\pi}{9}$ (dashed blue line), ~$\frac{\pi}{4}$ (dashed orange line)] when both Alice and Rob's qubits under decoherence.}
\end{figure}

Figure (\ref{ERP2})  shows time evolution of quantum entanglement
when the total two qubits system is coupled to the environment. It
shows that, compared with the case of only Rob's qubit undergoing
decoherence, the entanglement decreases more rapidly as the
acceleration increases. It is interesting to note that the sudden
death of entanglement appears at a finite time even for $r=0$, and a
lager acceleration also leads to an earlier appearance of the sudden
death as the  parameter $P$ increases.

In particular, when the acceleration approaches infinity, the sudden
death appears when $P\geq 1/2$, whereas it happens when
$P_R\geq\sqrt{2}/2$ when only Rob's qubit undergoes decoherence.
Thus, we come to the conclusion that the decoherence and loss of
entanglement generated by the Unruh effect will influence each other
in noninertial frames.

\section{summary}

\vspace*{0.5cm} In conclusion, we have found that, unlike the
isolated case in which the entanglement of Dirac fields survives
even in the limit of infinite acceleration \cite{Alsing-Mann}, the
entanglement could tend to zero for finite acceleration in this
system; and a lager acceleration leads to an earlier disappearance
of entanglement if either one or both subsystems experience a
decoherence. Thus, the decoherence and loss of entanglement
generated by the Unruh effect will influence each other remarkably
in noninertial frames. It is also shown that the sudden death of
entanglement will appear for any acceleration when both of the two
qubits interact with the environment. However, if only Rob's qubit
undergoes decoherence, the sudden death only takes place when the
acceleration parameter is greater than the ``critical point",
$r_c=\arcsin [(\sqrt{5}-1)/2]$. Our results can be applied to the
case in which Alice moves along a geodesic while Rob hovers near the
event horizon with an uniform acceleration and one or both of them
are in an amplitude-damping environment.

\vspace*{0.5cm} {\it Acknowledgments:} This work was supported by
the National Natural Science Foundation of China under Grant No
10875040; a key project of the National Natural Science Foundation
of China under Grant No 10935013; the National Basic Research of
China under Grant No. 2010CB833004, the Hunan Provincial Natural
Science Foundation of China under Grant No. 08JJ3010,  PCSIRT under
Grant No. IRT0964, and the Construct Program  of the National Key
Discipline.



\begin{thebibliography}{99}

\bibitem{Peres}
A. Peres and D. R. Terno, Rev. Mod. Phys. {\bf 76}, 93 (2004).

\bibitem{Boschi}
D. Boschi, S. Branca, F. De Martini, L. Hardy, and S. Popescu, Phys.
Rev. Lett. {\bf 80}, 1121 (1998).

\bibitem{Bouwmeester}
D. Bouwmeester, A. Ekert, and A. Zeilinger, \textit{The Physics of
Quantum Information} (Springer-Verlag, Berlin), 2000.

\bibitem{Bombelli-Callen}
L. Bombelli, R. K. Koul, J. Lee, and R. D. Sorkin, Phys. Rev. D {\bf
34}, 373 (1986).

\bibitem{Hawking-Terashima}
S. W. Hawking, Commun. Math. Phys. {\bf 43}, 199 (1975); Phys. Rev.
D {\bf 14}, 2460 (1976); H. Terashima, Phys. Rev. D {\bf 61}, 104016
(2000).

\bibitem{SRQIT1} A. Peres, P. F. Scudo, and D. R. Terno,
Phys. Rev. Lett. {\bf 88}, 230402 (2002).

\bibitem{Ging}
R. M. Gingrich and C. Adami Phys. Rev. Lett. {\bf 89}, 270402 (2002).

\bibitem{Alsing-Mann}
P. M. Alsing, I. Fuentes-Schuller, R. B. Mann, and T. E. Tessier,
Phys. Rev. A {\bf 74}, 032326 (2006).

\bibitem{Qiyuan}
Qiyuan Pan and Jiliang Jing, Phys. Rev. A {\bf77}, 024302 (2008);
Phys. Rev. D {\bf78}, 065015 (2008).

\bibitem{jieci1}
Jieci Wang, Qiyuan Pan, Songbai Chen, and Jiliang Jing, Phys. Lett.
B {\bf 677}, 186 (2009).

\bibitem{jieci2}
Jieci Wang, Junfeng Deng, and Jiliang Jing, Phys. Rev. A {\bf 81}, 052120  (2010).

\bibitem{Lamata}
L. Lamata, M. A. Martin-Delgado, and E. Solano, Phys. Rev. Lett.
{\bf 97}, 250502 (2006).

\bibitem{Schuller-Mann}
I. Fuentes-Schuller and R. B. Mann, Phys. Rev. Lett. {\bf 95},
120404 (2005).

\bibitem{Alsing-Milburn}
P. M. Alsing and G. J. Milburn, Phys. Rev. Lett. {\bf 91}, 180404
(2003).

\bibitem{Zurek}
W. H. Zurek, Rev. Mod. Phys. {\bf75}, 715 (2003).

\bibitem{Breuer}
H. P. Breuer and F. Petruccione, \textit{The Theory  of Open Quantum
Systems} (Oxford University Press, Oxford), 2002; H. Carmichael,
\textit{An Open Systems Approach to Quantum Optics} (Springer,
Berlin, 1993).

\bibitem{Giulini}
D. Giulini, E. Joos, C. Kiefer, J. Kupsch, I. O.  Stamatescu, and H.
D. Zeh, \textit{Decoherence and the Appearence of  a Classical World
in Quantum Theory} (Springer), 1996.

\bibitem{Schlosshauer}
 M. A. Schlosshauer, \textit{Decoherence and the Quantum-To-
Classical Transition} (Springer), 2007.

\bibitem{Brune}
M. Brune, E. Hagley, J. Dreyer, X. Maitre, A. Maali,  C. Wunderlich,
J. M. Raimond, and S. Haroche,  Phys. Rev. Lett. {\bf77}, 4887
(1996).

\bibitem{Myatt}
C. J. Myatt, B. E. King, Q. A. Turchette, C. A. Sackett, D.
Kielpinski, W. M. Itano, C. Monroe, and D. J. Wineland, Nature
{\bf403}, 269 (2000).

\bibitem{Salles}
A. Salles, F. de Melo1, M. P. Almeida1, M. Hor-Meyll, S. P. Walborn,
P. H. SoutoRibeiro, and L. Davidovich, Phys. Rev. A {\bf78}, 022322
(2008).

\bibitem{Brune1}
J. M. Raimond, M. Brune, and S. Haroche, Rev. Mod.  Phys. {\bf 73},
565 (2001).

\bibitem{Yu}
T. Yu and J. H. Eberly, Phys. Rev. Lett. {\bf97}, 140403 (2006).

\bibitem{Kraus}
K. Kraus, \textit{States, Effects and Operations: Fundamental
Notions of Quantum Theory} (Springer, Berlin), 1983.

\bibitem{Choi}
 M.-D. Choi, Linear Algebr Appl. {\bf 10}, 285 (1975).

\bibitem{Wootters}
W. K. Wootters, Phys. Rev. Lett. {\bf 80}, 2245 (1998).

\bibitem{Coffman}
V. Coffman, J. Kundu, and W. K. Wootters, Phys. Rev. A {\bf 61},
052306 (2000).

\end{thebibliography}
\end{document}